\documentclass[useAMS,usenatbib]{mn2e}

\newif\ifAMStwofonts
\usepackage{graphicx}
\usepackage{comment}

\title[A spectropolarimetric compendium of AGN]
      {A compendium of AGN inclinations with corresponding UV/optical continuum polarization measurements}
\author[F. Marin]
      {F.~Marin\thanks{frederic.marin@asu.cas.cz} \\
       Astronomical Institute of the Academy of Sciences, Bo{\v c}n\'{\i} II 1401, CZ-14100 Prague, Czech Republic}
\date{Accepted 2014 March 21. 
      Received 2014 March 21; 
      in original form 2013 December 12}

\pagerange{\pageref{firstpage}--\pageref{lastpage}}
\pubyear{2014}

\begin{document}

\maketitle

\label{firstpage}

%%%%%%%%%%%%%%%%%%%%%%%%%%%%%%%%%%%%%%%%%%%%%%%%%%%%%%%%%%%%%%%%%%%%%%%
\begin{abstract}
The anisotropic nature of active galactic nuclei (AGN) is thought to be responsible for the observational differences between 
type-1 (pole-on) and type-2 (edge-on) nearby Seyfert-like galaxies. In this picture, the detection of emission and/or
absorption features is directly correlated to the inclination of the system. The AGN structure can be further probed by
using the geometry-sensitive technique of polarimetry, yet the pairing between observed polarization and Seyfert type 
remains poorly examined. Based on archival data, I report here the first compilation of 53 estimated AGN inclinations matched 
with ultraviolet/optical continuum polarization measurements. Corrections, based on the polarization of broad emission lines, 
are applied to the sample of Seyfert-2 AGN to remove dilution by starburst light and derive information about the scattered
continuum alone. The resulting compendium agrees with past empirical results, i.e. type-1 AGN show low polarization degrees 
($P~\le$~1~\%) predominantly associated with a polarization position angle parallel to the projected radio axis of the system, 
while type-2 objects show stronger polarization percentages ($P~>$~7~\%) with perpendicular polarization angles. The transition 
between type-1 and type-2 inclination occurs between 45$^\circ$ and 60$^\circ$ without noticeable impact on $P$. The compendium 
is further used as a test to investigate the relevance of four AGN models. While an AGN model with fragmented regions matches 
observations better than uniform models, a structure with a failed dusty wind along the equator and disc-born, ionized, polar 
outflows is by far closer to observations. However, although the models correctly reproduce the observed dichotomy between 
parallel and perpendicular polarization, as well as correct polarization percentages at type-2 inclinations, further work is
needed to account for some highly polarized type-1 AGN
\end{abstract}


%%%%%%%%%%%%%%%%%%%%%%%%%%%%%%%%%%%%%%%%%%%%%%%%%%%%%%%%%%%%%%%%%%%%%%%
\section{Introduction}
\label{Intro}

The unified model of active galactic nuclei (AGN; \citealt{Antonucci1993}) states that most of the observational 
differences between type-1 and type-2 Seyfert-like galaxies arise from an orientation effect. According to this theory, 
the disappearance of ultraviolet (UV) and optical broad emission features at type-2 inclinations can be explained 
by the presence of an obscuring, circumnuclear material along the equatorial plane of the AGN (the so-called dusty torus)
that hides both the central engine and the photoionized broad line regions (BLRs; the low ionization line LIL BLR and
the highly ionized line HIL BLR; \citealt{Rowan1977,Osterbrock1978}). A type-2 viewing angle can then be defined as a 
line of sight towards the central source that intercepts the equatorial dust, while type-1 inclination allows a direct view 
of the central engine. The observational lack of type-1 AGN with edge-on host galaxy \citep{Keel1980,Lawrence1982} suggests that 
dust along Seyfert~1 galaxy discs may obscure the HIL and LIL BLR and make the AGN appear like Seyfert~2s \citep{Maiolino1995}. 
The number count of Seyfert~1 objects is thus expected to be small, even if the fraction of type-1 against type-2 AGN in the 
nearby Universe still needs to be properly determined. Estimating the orientation of a large sample of AGN is necessary to 
verify the assumptions of the unified model, and check whether all the differences between Seyfert~1 and Seyfert~2 objects 
can be explained by inclination or if morphological differences must also be taken into account.

In this regard, polarization has proven to be a solid tool to investigate the inner structure of AGN. The spectropolarimetric
measurements of NGC~1068 by \citet{Miller1983} helped to identify electron and dust scattering as the main 
mechanisms producing a continuum polarization in radio-quiet AGN. Going further, the extensive high-resolution, high 
signal-to-noise spectropolarimetric observation of the same AGN by \citet{Antonucci1985} revealed the presence of
highly polarized, broad, symmetric Balmer and permitted Fe~{\sc ii} lines. The polarization spectrum was found to be closely similar 
to typical Seyfert~1 galaxies, supporting the idea that Seyfert~2 AGN are hiding Seyfert~1 core behind the dusty torus. This 
discovery was a key argument in favour of a unified model of AGN. 

Spectropolarimetry is thus a powerful method to probe the validity of any AGN model, as the computed fluxes shall match both 
observational intensity, polarization percentage and polarization angle, reducing the number of free parameters/degeneracies
\citep{Kartje1995,Young2000,Goosmann2007,Marin2012a}. In order to model a peculiar source, the observer's viewing angle $i$ has 
to be set (e.g. $i \sim$ 70$^\circ$ for NGC~1068; \citealt{Honig2007,Raban2009}) to explore the resulting polarization
\citep{Goosmann2011,Marin2012c}. The impact of the system's orientation on to the net polarization can lead to significantly 
different results, especially when the observer's line of sight matches the half-opening angle of the obscuring region 
\citep{Marin2012a}. To be consistent with observation, an investigation of the model over a broad range of inclination
must be undertaken. However, any comparison between the observed polarization and the theoretical orientation of individual AGN
is hampered by the lack of a data base that combines inclination and polarization. 

It is the aim of this paper to provide the first spectropolarimetric compendium of Seyfert-like galaxies, gathering 
observed continuum polarizations from literature correlated with estimated inclinations of individual AGN. In Sect.~\ref{Comp}, I 
investigate the different observational and numerical techniques used to estimate the inclination of 53 objects, and match 
the sources with their archival UV/optical polarization measurement, whenever it is feasible. To illustrate the significance 
of this catalogue when comparing models to observations, in Sect.~\ref{Analysis}, I pick four different, competitive AGN models 
from the literature and analyse them in the framework of this compendium. In Sect.~\ref{Discussion}, I review the 
successes and potential improvements of AGN modelling, explore the problematic, high polarization levels of peculiar type-1 
objects and discuss possible bias on the estimation of inclination. Finally, conclusions are drawn in Sect.~\ref{Conclusion}.

%%%%%%%%%%%%%%%%%%%%%%%%%%%%%%%%%%%%%%%%%%%%%%%%%%%%%%%%%%%%%%%%%%%%%%%
\section{The compendium}
\label{Comp}

While polarimetric measurement of AGN are difficult to obtain due to intrinsically low polarization degrees in type-1 objects 
\citep{Berriman1989,Berriman1990,Smith2002a} and flux dilution by unpolarized starlight in type-2s \citep{Antonucci2002a}, it is 
more straightforward to measure polarization than to estimate the inclination of the system, as orientation is not easily derived directly 
from observations. Hence, the following sections present different techniques to infer the orientation of AGN (with potential caveats 
discussed in Sect.~\ref{Discussion:Bias}), the selection criteria used to select/remove estimated inclinations, corrections that have
to be made to most of the Seyfert-2 AGN and the final compilation of data. For the remaining of this paper, the terminology 
``inclination of the system'' will refer to the nuclear, not the host, inclination.

\subsection{Estimation of the system's inclination}
\label{Comp:Incl}

{\sc method A}. Based on the tight correlation found by \citet{Gebhardt2000}, \citet{Ferrrarese2000} and \citet{Merritt2001} between the mass of 
the central supermassive black hole and the bulge velocity dispersion in nearby galaxies, \citet{Wu2001} and \citet{Zhang2002} developed 
a method to derive the inclination angle of nearby Seyfert~1 AGN. By assuming a Keplerian motion of the LIL BLR and a similar mass/velocity 
dispersion between type-1 AGN and regular galaxies, they estimated the orientation angles $i$ for a variety of Seyfert~1 objects, with associated 
errors calculated from the uncertainties of both the black hole mass (obtained by reverberation mapping techniques; \citealt{Blandford1982})
and the measured velocity dispersion.

{\sc method B}. The inclination estimation of the Seyfert~1 galaxy ESO~323-G077 comes from the optical spectropolarimetric measurement 
achieved by \citet{Schmid2003}, who detected very high levels of linear polarization (up to 7.5~\% at 3600~\AA). Within the framework of 
the unified model, those levels are inconsistent with the polarization degrees produced by an object seen in the polar orientation 
\citep{Marin2012a}. \citet{Schmid2003} argued that the system must be partially hidden by the dusty torus and tilted by $\sim$~45$^\circ$ 
with respect to the observer's line of sight to produce such a high polarization degree. The same method was previously applied to 
Fairall~51 (continuum polarization 4.12~\% $\pm$ 0.03~\%) by \citet{Schmid2001}, who also derived an inclination of $\sim$~45$^\circ$.

{\sc method C}. An increasing amount of X-ray bright, type-1 AGN shows an asymmetrically blurred emission feature at 6.4~keV, associated with 
iron fluorescence in near-neutral material \citep{Reeves2006}. Interestingly, the line broadening caused by Doppler effects and gravitational 
plus transverse redshifts can be used to numerically probe the inclination of the system \citep{Fabian1989}. In \citet{Nandra1997}, 
this characteristic line profile is equally fitted within a Schwarzschild or a Kerr metric (even if recent modelling seems to favour maximally 
rotating black holes in the centre of type-1 AGN; \citealt{Bambi2011,Bambi2013}), giving a mean Seyfert~1 inclination of 30$^\circ$.

{\sc method D}. Constraints on the inclination of NGC~1097 are derived by \citet{Storchi1997}, who applied an eccentric accretion ring model 
to the observed broad, double-peaked Balmer emission lines. Between 1991 and 1996, the double-peaked H$\alpha$ line of NGC~1097 evolved from a 
red-peak dominance \citep{Storchi1993} to a nearly symmetrical profile \citep{Storchi1995} and up to a blue-peak dominance \citep{Storchi1997}. 
This line profile evolution can be explained by a refinement of the precessing, planar, elliptical accretion-ring model developed by 
\citet{Storchi1995} and \citet{Eracleous1995}, to fit the data using an eccentric accretion disc inclined by 34$^\circ$.

{\sc method E}. \citet{Hicks2008} measured the two-dimensional distribution and kinematics of the molecular, ionized, and highly ionized gas in 
the inner regions of a sample of radio-quiet, type-1 AGN using high spatial resolution, near-infrared (IR) spectroscopy. Based on a model developed
by \citet{Macchetto1997}, they assumed that a gravitational well, created by the combined action of a central supermassive black hole and 
a distant stellar population, is driving the circular motion of a coplanar thin disc, reproducing the observed emission line gas kinematics.
Exploring four free parameters (disc inclination, position angle of its major axis, black hole mass and mass-to-light ratio), \citet{Hicks2008} 
statistically estimated the inclination of NGC~3227, NGC~4151 and NGC~7469 using a Bootstrap Gaussian fit \citep{Efron1979}.

{\sc method F}. To determine the inclination of a sample of nearby AGN, \citet{Fischer2013} explored the three-dimensional geometry and kinematics of 
the narrow-line regions (NLRs) of AGN, observing both type-1 and type-2 objects. Resolved by [O~{\sc iii}] imaging and long-slit spectroscopy, most of 
the AGN show a bi-conical structure which can be morphologically and kinematically constrained. However, to extract information about the AGN 
orientation, a kinematic model must be generated. Using uniform, hollow, bi-conical models with sharp edges, \citet{Fischer2013} were able to 
statistically derive a set of morphological parameters (including orientation) for 17 objects out of 53.

{\sc method G}. Relatively bright and situated in the nearby Universe, NGC~1068 is an archetypal Seyfert~2 galaxy, observed during the last fifty years.
Taking advantage of past near and mid-IR photometric and interferometric observations \citep{Jaffe2004,Wittkowski2004}, \citet{Honig2007}
applied a three-dimensional radiative transfer code to a clumpy, dusty structure in order to reproduce the observed spectral energy distribution (SED).
Among new constraints on the bolometric luminosity and the IR optical depth of the torus, \citet{Honig2007} estimated the overall inclination of 
NGC~1068 to be close to 70$^\circ$.

{\sc method H}. \citet{Borguet2010} examined the generation of C~{\sc iv} line profiles in broad absorption line (BAL) quasars using a two-component wind model.
By modelling a structure based on stellar wind laws and composed of axisymmetric, polar and equatorial outflows filled with 2-level atoms, they 
succeeded to reproduce a large set of BAL profiles and concluded that the viewing angle to the wind is generally large. Unfortunately, degeneracies
in line profile fitting do not allow stronger constraints.

{\sc method I}. Finally, \citet{Wills1992} investigated their own polarimetric and photometric observations of the type-2 quasar IRAS~13349+2438 
in the context of an axisymmetric distribution of scatterers to explain the alignment of polarization with the major axis of the host galaxy.
Using a model of a dusty disc parallel to the plane of the galaxy, similar to a usual dusty torus, surrounding the continuum source and the LIL BLR, 
they showed that both the observed polarization in the continuum and in the broad H$\alpha$ line could be reproduced if the inclination of the observer 
is about 52$^\circ$ with respect to the symmetry axis of their model.

\subsection{Selection criteria}
\label{Comp:Tables}

Once inclinations are obtained, I match them with UV/optical spectropolarimetric measurements, whenever it was possible. The methods presented in 
Sect.~\ref{Comp:Incl} derive about 100 AGN orientations but only 53 of them have published continuum polarization measurements. Moreover, not all 
of the estimated inclinations are unique and a selection has to be done whenever two methods, or more, give different estimations for the 
same AGN. Tab.~\ref{Table:Reject} lists discrepancies of duplicate inclinations. Reasons for the choice of a given inclination are discussed below.

\begin{table*}
  \centering
  {   
    \begin{tabular}{|c|c|c|c|}
      \hline
      {\bf Object} 	& {\bf Type} 	& {\bf $i_{\rm sel}$ ($^\circ$)} 		& {\bf $i_{\rm rej}$ ($^\circ$)} \\
      \hline
      3C~120 		& 1 		& 22.0$^{+9.3}_{-7.7}$ \citep{Wu2001} 		& 88$^{+2}_{-1}$ \citep{Nandra1997} \\
      Fairall~9		& 1 		& 35.0 \citep{Zhang2002}			& 89$^{+1}_{-49}$ \citep{Nandra1997} \\
      IC~4329A		& 1 		& 10$^{+13.0}_{-10.0}$ \citep{Nandra1997}	& 5.0 \citep{Zhang2002} \\
      Mrk~279		& 1 		& 35.0 \citep{Fischer2013}			& 13.0 \citep{Zhang2002} \\          
      Mrk~509		& 1 		& 19.0 \citep{Zhang2002}			& 89$^{+1}_{-89}$ \citep{Nandra1997} \\   
      NGC~3227		& 1 		& 14.2 $\pm$ 2.5 \citep{Hicks2008}		& 15.0 \citep{Fischer2013} \\ 
      ~			& ~		& ~						& 21$^{+7}_{-21}$ \citep{Nandra1997}\\
      ~			& ~		& ~						& 37.5$^{+17.3}_{-25.4}$ \citep{Wu2001}\\    
      NGC~3516		& 1 		& 26$^{+3}_{-4}$ \citep{Nandra1997}		& 38.3 $\pm$ 7.6 \citep{Wu2001} \\    
      NGC~3783		& 1 		& 15.0 \citep{Fischer2013}			& 40$^{+12}_{-40}$ \citep{Nandra1997}\\
      ~			& ~		& ~						& 38.0 \citep{Zhang2002} \\   
      NGC~4051		& 1 		& 19.6$^{+10.4}_{-6.6}$ \citep{Wu2001}		& 25$^{+12}_{-4}$ \citep{Nandra1997}\\
      ~			& ~		& ~						& 10.0 \citep{Fischer2013} \\   
      NGC~4151		& 1 		& 9$^{+18}_{-9}$ \citep{Nandra1997}		& 19.8 $\pm$ 2.9 \citep{Hicks2008}\\
      ~			& ~		& ~						& 45 \citep{Fischer2013}\\
      ~			& ~		& ~						& 60$^{+30}_{-30.6}$ \citep{Wu2001} \\   
      NGC~5548		& 1 		& 47.3$^{+7.6}_{-6.9}$ \citep{Wu2001}		& 10$^{+80}_{-10}$ \citep{Nandra1997} \\   
      NGC~7469		& 1 		& 15.0 $\pm$ 1.8 \citep{Hicks2008}		& 20$^{+70}_{-20}$ \citep{Nandra1997}\\
      ~			& ~		& ~						& 13.0 \citep{Zhang2002} \\  
      NGC~1068		& 2 		& 70.0 \citep{Honig2007}			& 85.0 \citep{Fischer2013} \\        	
      \hline
    \end{tabular}
  }
  \caption{Selected ($i_{\rm sel}$) and rejected ($i_{\rm rej}$) nuclear inclinations according to the 
	    selection criteria presented in Sect.~\ref{Comp:Incl}.}
  \label{Table:Reject}
\end{table*}

~\

{\sc 3c~120}. The inclination of 3C~120 found by \citet{Nandra1997}, $i$~=~88${^\circ}^{+2}_{-1}$, is rejected in favour of the estimation by 
\citet{Wu2001}, who found $i$~=~21${^\circ}^{+9.3}_{-7.7}$, a result in a better agreement with the type-1 classification of 3C~120\footnote{3C~120 
is a type-1, broad-line, X-ray bright radio galaxy showing an episodic superluminal jet outflow \citep{Marscher2002}. 3C~120 is sometimes 
included in radio-quiet AGN surveys as its X-ray spectrum shows a strong relativistic iron K$\alpha$ emission \citep{Nandra1997}.}.

{\sc fairall~9}. Due to the huge error bars derived by \citet{Nandra1997} on the orientation of Fairall~9 (89${^\circ}^{+1}_{-49}$), covering the full 
permitted range of inclination for a type-2 object plus a fraction of the permitted range of type 1s, the viewing angle computed by 
\citet{Zhang2002} is favored.

{\sc ic~4329a}. The inclination derived by \citet{Zhang2002} in the case of IC~4329A (5.0$^\circ$) is compatible within the error bars of the 
estimation computed by \citet{Nandra1997}. The latest (10${^\circ}^{+13.0}_{-10.0}$) is thus selected in order to concur with the two values.

{\sc mrk~279}. Estimated inclinations of Mrk~279 are rather different between \citet{Zhang2002}, $i$~=~13.0$^\circ$, and \citet{Fischer2013}, 
$i$~=~35.0$^\circ$, especially since they do not have overlapping error bars. The inclination derived by \citet{Fischer2013} is favoured as 
\citet{Zhang2002} were not able to recover the measured stellar velocity dispersion of Mrk~279 and had to estimate it from the [O {\sc iii}] emission 
line, introducing another potential bias in their final AGN orientation. 

{\sc mrk~509}. The inclination estimated by \citet{Nandra1997} covers the whole range of inclination possible for an AGN (89${^\circ}^{+1}_{-89}$)
and therefore does not make much sense. The inclination of Mrk~509 by \citet{Zhang2002} is thus selected.

{\sc ngc~3227}. There are four different estimations for the viewing angle of NGC~3227. Two have very large error bars \citep{Nandra1997,Wu2001}
overlapping the two others estimates by \citet{Hicks2008} and \citet{Fischer2013}. Since the two later inclinations are very similar but from 
totally different estimation methods, they are likely to be representative of the real inclination of NGC~3227. Hence I favour the one of 
\citet{Hicks2008}, 14.2$^\circ$ $\pm$ 2.5$^\circ$, which agrees with the value found by \citet{Fischer2013}, 
15.0$^\circ$.

{\sc ngc~3516}. The values estimated by \citet{Nandra1997}, $i$~=~26${^\circ}^{+3}_{-4}$, and \citet{Wu2001}, $i$~=~38.3$^\circ$ $\pm$ 7.6$^\circ$, 
are not overlapping but still very close to each other. However, provided that \citet{Wu2001} had to artificially estimate the errors on the 
black hole mass for NGC~3516 while \citet{Nandra1997} derived it from their simulation, the estimation made by \citet{Nandra1997} is used.

{\sc ngc~3783}. Similarly to the case of Mrk~279, estimations of the orientation of NGC~3783 are rather different between \citet{Zhang2002} and 
\citet{Fischer2013} and, for the same reasons, the value derived by \citet{Fischer2013} is selected. The inclination recovered by \citet{Nandra1997} 
for NGC~3783, covering the whole possible inclination range for type-1 objects, is discarded.

{\sc ngc~4051}. Both estimations made by \citet{Nandra1997}, $i$~=~25${^\circ}^{+12}_{-4}$, and \citet{Wu2001}, $i$~=~19.6${^\circ}^{+10.4}_{-6.6}$, 
are compatible and their error bars nearly fully overlap. The viewing angle taken from \citet{Wu2001} having slightly higher error bars, this value 
is chosen to fully concur with the estimations from \citet{Nandra1997} and to be representative of the value derived by \citet{Fischer2013}, 10.0$^\circ$.

{\sc ngc~4151}. The inclination angle of NGC~4151 derived by \citet{Wu2001} is not considered due to its huge error bars that cover two thirds of 
the possible AGN inclinations (60${^\circ}^{+30}_{-30.6}$). The overlapping values, from different methods, found by \citet{Nandra1997}, 9${^\circ}^{+18}_{-9}$, 
and \citet{Hicks2008}, 19.8$^\circ$ $\pm$ 2.9$^\circ$, exclude the fourth one derived by \citet{Fischer2013}, 45.0$^\circ$. Finally, the estimation made 
by \citet{Nandra1997} is preferred as it fully covers the potential inclination calculated by \citet{Hicks2008}.

{\sc ngc~5548}. Similar to the case of Mrk~509, the inclination of NGC~5548 evaluated by \citet{Nandra1997}, $i$~=~10${^\circ}^{+80}_{-10}$, 
is rejected in favour of the one derived by \citet{Wu2001}, $i$~=~47.3${^\circ}^{+7.6}_{-6.9}$.

{\sc ngc~7469}. The viewing angle derived by \citet{Zhang2002} is consistent with the one derived by \citet{Hicks2008}, within the error
bars of the later. The inclination of NGC~7469 estimated by \citet{Nandra1997} is not worth considering due to its huge error bars
(20${^\circ}^{+70}_{-20}$).

{\sc ngc~1068}. Finally, the estimations of \citet{Fischer2013} and \citet{Honig2007} are different and do not overlap. However, the inclination
derived by \citet{Honig2007} is supported by the three-dimensional structure of the nuclear region of NGC~1068 reconstructed by 
\citet{Kishimoto1999} using a different set of observations. As \citet{Kishimoto1999} derived a similar, $\sim$~70$^\circ$, inclination, the 
estimation of \citet{Honig2007} is thus favoured.

\subsection{Revisited polarization of Seyfert~2s}
\label{Comp:Seyfert2}

Similarly to the inclination estimates in the previous section, I have found for a very few number of objects several continuum polarization 
measurements achieved by different authors. The level of polarization and the polarization position angle (measured from north through east) were 
coherent between the different observing campaigns, regardless of the epoch. I have thus favoured polarimetric measurement from Seyfert atlases 
\citep{Martin1983,Brindle1990,Kay1994,Ogle1999,Smith2002a}. 

However, not all of these measurements are reliable. Most, if not all, type-2 AGN are dominated by relatively large, unpolarized starlight fluxes 
\citep{Antonucci2002b}. Removing the contribution from old stellar populations drastically increases the resulting continuum polarization but previous 
polarimetric type-2 atlases still recorded low, 1 -- 3~\%, polarization degrees \citep{Martin1983,Kay1994,Smith2002b}. Such behaviour is in disagreement 
with the unified model, where radiation escapes from the inner parts of the obscuring equatorial torus by perpendicular scattering into the polar outflows, 
carrying a large amount of polarization \citep{Antonucci1993}. High polarization percentages are thus expected, but not observed, for type-2 objects 
\citep{Miller1990}. \citet{Miller1990} reported that even after starlight subtraction, the remaining continuum is dominated by another component 
responsible for dilution of the polarized flux, now identified as originating from starburst regions. To estimate the continuum polarization of a 
given object after corrections for the interstellar polarization and dilution by host galaxy starlight, \citet{Tran1995a,Tran1995b,Tran1995c} proposed 
to measure the equivalent width (EW) of the broad Balmer lines in both intensity and polarized flux spectra, since polarized flux spectra have the advantage 
to suppress low polarization emission from starlight and narrow emission lines. If no additional unpolarized, or very little polarized, continuum 
superimposes on the polarization originating from the scattered light alone, the intrinsic polarization in the line and the adjacent continuum should 
be equal. This is the case for NGC~1068, where the broad lines have the same polarization as the continuum \citep{Antonucci1994}. Unfortunately, nearly 
all the remaining measurements of type-2 AGN polarization are likely to be biased downward. 

In the following I revisit and, when necessary and possible, revise the estimated continuum polarization of type-2 objects to be included in the compendium.
The best way to estimate the continuum polarization due only to scattered light is to divide the polarized flux by the total flux across the broad emission lines, 
as suggested by \citet{Tran1995c} and \citet{Antonucci2002a}. Only the broad lines polarization is a reliable indicator of the polarization of the scattered 
component. However, by definition, broad lines are not detected in the total flux spectra of type-2s. To overcome this problem, I compare the typical 
EW of type-1 polarized, broad, Balmer emission lines H$\alpha$~$\lambda$6563 (EW~$\approx$~400~\AA; \citealt{Smith2002a}) and 
H$\beta$~$\lambda$4861 (EW~$\approx$~80~\AA; \citealt{Young1997}) to the EW of the polarized H$\alpha$~$\lambda$6563 and H$\beta$~$\lambda$4861 lines of
the compendium-selected Seyfert 2s. This is a first-order approximation, but it is justified by the fact that the broad lines have the same EW in polarized flux in type-1 
and type-2 AGN \citep{Antonucci2002b}. If the ratio is equal to unity, the continuum polarization is correct; if larger than unity, the continuum must be 
revised by the same factor. In the case where no spectra are available to estimate the EW of broad emission lines in polarized flux, the continuum 
polarization reported by previous Seyfert atlases will be used as a lower limit \citep{Goodrich1994}.

~\

I revisit the polarization of the Circinus galaxy measured by \citet{Alexander2000}, 1.9~\%, using the only broad emission line detected 
in its polarized flux spectrum, namely H$\alpha$~$\lambda$6563 (H$\beta$~$\lambda$4861 being only marginally detected, $\sim$~2$\sigma$). The EW, 
estimated using a Lorentzian profile \citep{Kollatschny2012}, is about 34~\AA. From then, the ratio between type-1 H$\alpha$ EW ($\approx$~400~\AA)
and EW$_{\rm H\alpha~Circinus}$ is equal to 11.8 and the resulting continuum polarization is equal to 22.4~\%. Interestingly, \citet{Oliva1998} 
have derived a similar result (25~\%) using a simple, numerical model applied to their own spectropolarimetric measurement of the Circinus galaxy.
 
Mrk~3 has been observed by \citet{Tran1995a}, who found after correction for interstellar polarization and dilution by starlight a continuum 
polarization of 7.0~\%. Using H$\alpha$~$\lambda$6563 and H$\beta$~$\lambda$4861 as diagnostic lines from \citet{Tran1995b}, the EW of H$\alpha$ 
is estimated to be $\sim$~360~\AA~and EW H$\beta$ $\sim$~65~\AA. The ratio are 1.11 and 1.23, respectively. The scattered light is thus expected 
to have an intrinsic polarization of 7.77 -- 8.61~\%. 
   
The small bump around 6563~\AA~in the percentage of polarization spectrum of NGC~1667 was attributed to the red wing of the H$\alpha$ + 
[N {\sc ii}] profile by \citet{Barth1999}, who consequently stated that no polarized H$\alpha$ emission have been detected. However, the polarization
spectra are dominated by high noise levels, probably hiding the appearance of the broad wings of the line. The EW of the polarized H$\alpha$~$\lambda$6563
line is consequently very similar to the EW measured in the total flux spectrum, i.e. 14.3~\AA. The resulting ratio, 28.0, can be used to account for 
an upper limit of the intrinsic continuum polarization, which is then set between 0.35~\% \citep{Barth1999} and 9.8~\%.
 
Both H$\alpha$~$\lambda$6563 and H$\beta$~$\lambda$4861 polarized broad lines can be detected in the polarized flux spectrum of NGC~4507 
\citep{Moran2000}. The estimated EW are 165~\AA~(H$\alpha$) and 30~\AA~(H$\beta$), raising the continuum polarization from its initial 
value (6.1~\%; \citealt{Moran2000}) to 14.8 -- 16.3~\%.

NGC~5506 has been observed by \citet{Kay1994}, who found a continuum polarization of 2.60~\% $\pm$ 0.41~\%. From the spectropolarimetric measurements 
of a sample of nearby Compton-thin ($N_{\rm H}~<$~10$^{23}$~cm$^{-2}$) Seyfert 2s, \citet{Lumsden2004} found no evidence for a broad H$\alpha$~$\lambda$6563 
line in NGC~5506. They argued that the cause of non-detection of the broad lines in the polarized spectrum of this AGN can be due to an extended 
obscuring region rather than non-existence. I estimate an upper limit on the EW for the polarized H$\alpha$~$\lambda$6563 line ($\sim$~10~\AA)  
looking at the total flux spectrum and set the corrected polarization degree between 2.6~\% and 100~\%.
  
\citet{Tran1995b} measured the continuum polarization of NGC~7674 to be 3.8~\% after removing the starlight contribution. From their 
spectra, I estimate the EW of the H$\alpha$~$\lambda$6563 and H$\beta$~$\lambda$4861 lines to be 208~\AA~and 40~\AA, respectively. 
The ratio are 1.72 and 2.0, increasing the intrinsic polarization of the scattered light to 6.54 -- 7.6~\%. 

NGC~1068 has proven to be remarkable in a sense that the polarization degree of its broad emission lines is similar to that of the continuum.
This peculiar feature is also shared by the broad H$\alpha$ line/continuum polarization in IRAS~13349+2438 and no revisions are necessary for 
these two objects. The high level of polarization detected in Mrk~78 by \citet{Kay1994}, 21.0~\% $\pm$ 9.0~\%, is significantly different from 
the rest of the sample and might face the same physical conditions, if not suffering from an upward bias (since polarization is a positive-definite 
quantity). The absence of spectra make it impossible to verify this assumption.
 
The polarization spectra of the seven BAL quasars extracted from \citet{Ogle1999}, 0019+0107, 0145+0416, 0226--1024, 0842+3431, 
1235+1453, 1333+2840 and 1413+1143, are too noisy to measure the EW of isolated broad emission lines in polarized flux. The few broad emission 
lines detected in total flux are either merged with some other lines (L$\alpha$ + N~V; Al~III + C~III]), truncated or poorly resolved. 
Most of them are not even detected in polarized flux. The reported polarization continuum will then be used as lower limits.
 
Unfortunately, there are no detailed polarization spectra published so far for Mrk~34, Mrk~573, Mrk~1066 and NGC~5643. 

~\

Over the 20 type-2 AGN, 3 of them did not need any correction, 6 were revised and 11 can only be used as basic lower limits for the intrinsic polarization
of the scattered light. Type-1 AGN, once corrected for starlight contribution, do not suffer from an additional dilution component and are thus directly
exploitable. The detailed lists of the AGN sampled, with inclination matched to continuum polarization, are given in Tab.~\ref{Table:Type1} and 
Tab.~\ref{Table:Type2}, for type-1 and type-2 objects respectively.

\begin{table*}
  \centering
  {   
   \begin{tabular}{|c|c|c|c|c|c|c|c|}
   \hline {\bf Object}	& {\bf Waveband (\AA)}	& {\bf Pol. degree (\%)}	& {\bf Pol. angle ($^\circ$)}	& {\bf Ref.}	& {\bf Inclination ($^\circ$)}	& {\bf Ref.}	& {\bf Method}\\
   \hline 3C~120	& 3800 -- 5600 		& 0.92 $\pm$ 0.25 		& 103.5 $\pm$ 7.9		& Mar83		& 22.0$^{+9.3}_{-7.7}$		& Wu01		& A\\
   	  Akn~120	& 3800 -- 5600 		& 0.65 $\pm$ 0.13		& 78.6 $\pm$ 5.7		& Mar83		& 42.0				& Zha02		& A\\
	  Akn~564	& 6000 -- 7500 	 	& 0.52 $\pm$ 0.02		& 87.0 $\pm$ 1.3		& Smi02		& 26.0				& Zha02		& A\\
   	  ESO~323-G077	& 3600 			& 7.5 				& 84				& Sch03		& 45.0				& Sch03		& B\\
	  Fairall~9	& 3800 -- 5600 		& 0.4 $\pm$ 0.11 		& 2.4 $\pm$ 7.6			& Mar83		& 35.0				& Zha02		& A\\
	  Fairall~51	& 4700 -- 7200 		& 4.12 $\pm$ 0.03 		& 141.2 $\pm$ 0.2		& Smi02		& 45.0				& Sch01		& B\\
	  IC~4329A	& 5000 -- 5800 		& 5.80 $\pm$ 0.26		& 42.0 $\pm$ 1.0		& Bri90		& 10$^{+13.0}_{-10.0}$ 		& Nan97		& C\\
	  PG~1211+143	& 4700 -- 7200 	 	& 0.27 $\pm$ 0.04		& 137.7 $\pm$ 4.5		& Smi02		& 31.0				& Zha02		& A\\
	  MCG-6-30-15	& 5000 -- 5800 		& 4.06 $\pm$ 0.45		& 120.0 $\pm$ 3.0		& Bri90		& 34.0$^{+5.0}_{-6.0}$ 		& Nan97		& C\\
	  Mrk~79	& 3800 -- 5600 		& 0.34 $\pm$ 0.19		& 0.4 $\pm$ 16.2		& Mar83		& 58.0				& Zha02		& A\\
	  Mrk~110	& 3200 -- 8600 		& 0.17 $\pm$ 0.08		& 18.0 $\pm$ 15.0		& Ber90		& 37.4$^{+9.2}_{-9.5}$		& Wu01		& A\\
	  Mrk~279	& 6000 -- 7500 		& 0.48 $\pm$ 0.04		& 58.9 $\pm$ 2.4		& Smi02		& 35.0				& Fis13		& F\\
	  Mrk~335	& 3800 -- 5600 		& 0.48 $\pm$ 0.11		& 107.6 $\pm$ 6.9		& Mar83		& 20.0				& Zha02		& A\\
	  Mrk~478	& 3800 -- 5600 	 	& 0.46 $\pm$ 0.15		& 44.9 $\pm$ 9.5		& Mar83		& 25.0				& Zha02		& A\\
	  Mrk~486	& 3800 -- 5600 	 	& 3.40 $\pm$ 0.14		& 136.8 $\pm$ 1.2		& Mar83		& 16.0				& Zha02		& A\\
	  Mrk~509	& 3800 -- 5600 		& 1.09 $\pm$ 0.15		& 146.5 $\pm$ 4.0		& Mar83		& 19.0				& Zha02		& A\\
	  Mrk~590	& 3800 -- 5600	 	& 0.32 $\pm$ 0.30		& 105.9 $\pm$ 26.6		& Mar83		& 17.8$^{+6.1}_{-5.9}$		& Wu01		& A\\
	  Mrk~705	& 4700 -- 7200 	 	& 0.46 $\pm$ 0.07		& 49.3 $\pm$ 6.5		& Smi02		& 16.0				& Zha02		& A\\
	  Mrk~707	& 3800 -- 5600 	 	& 0.20 $\pm$ 0.24		& 140.9 $\pm$ 52.0		& Mar83		& 15.0				& Zha02		& A\\
	  Mrk~766	& 4500 -- 7100 		& 3.10 $\pm$ 0.80		& 90.0				& Bat11 	& 36.0$^{+8.0}_{-7.0}$		& Nan97		& C\\
	  Mrk~841	& 4500 -- 7500	 	& 1.00 $\pm$ 0.03		& 103.4 $\pm$ 1.0		& Smi02		& 26.0$^{+8.0}_{-5.0}$		& Nan97		& C\\
	  Mrk~896	& 3800 -- 5600 	 	& 0.55 $\pm$ 0.13		& 1.9 $\pm$ 7.1			& Mar83		& 15.0				& Zha02		& A\\
	  Mrk~1239	& 3800 -- 5600 	 	& 4.09 $\pm$ 0.14		& 136.0 $\pm$ 1.0		& Mar83		& 7.0				& Zha02		& A\\
	  NGC~1097	& 5100 -- 6100 		& 0.26 $\pm$ 0.02		& 178 $\pm$ 2.0			& Bar99		& 34.0 	 			& Sto97		& D\\
	  NGC~1365	& 5000 -- 5900 		& 0.91 $\pm$ 0.18		& 157 $\pm$ 6.0			& Bri90		& 57.5 $\pm$ 2.5 	 	& Risa13	& C\\
	  NGC~3227	& 5000 			& 1.3 $\pm$ 0.1			& 133 $\pm$ 3.0			& Sch85		& 14.2 $\pm$ 2.5		& Hic08		& E\\
	  NGC~3516	& 4500 -- 7500		& 0.15 $\pm$ 0.04		& 30.1 $\pm$ 8.0		& Smi02		& 26$^{+3}_{-4}$		& Nan97		& C\\
	  NGC~3783	& 4500 -- 7500 		& 0.52 $\pm$ 0.02		& 135.5 $\pm$ 1.0		& Smi02		& 15.0				& Fis13		& F\\
	  NGC~4051	& 4500 -- 7500 		& 0.55 $\pm$ 0.04		& 82.8 $\pm$ 1.8		& Smi02		& 19.6$^{+10.4}_{-6.6}$		& Wu01		& A\\
	  NGC~4151	& 4600 -- 7400		& 0.26 $\pm$ 0.08     		& 62.8 $\pm$ 8.4		& Mar83		& 9$^{+18}_{-9}$ 		& Nan97		& C\\
	  NGC~4593	& 6000 -- 7600 		& 0.14 $\pm$ 0.05		& 109.5 $\pm$ 10.8		& Smi02		& 21.6 $\pm$ 10.5		& Wu01		& A\\
	  NGC~5548	& 6000 -- 7500 		& 0.69 $\pm$ 0.01		& 33.2 $\pm$ 0.5		& Smi02		& 47.3$^{+7.6}_{-6.9}$		& Wu01		& A\\
	  NGC~7469	& 6000 -- 7500 		& 0.18 $\pm$ 0.01		& 76.8 $\pm$ 1.7		& Smi02		& 15.0 $\pm$ 1.8 		& Hic08		& E\\
   \hline
   \end{tabular}
  }
  \caption{Recorded average continuum polarization states and inclinations of 33 type-1 AGN. 
	    The first reference column is related to polarization measurements, the second to estimations
	    of the orientation. Methods used to determine the inclination of the system are described in Sect.~\ref{Comp:Incl}.
	    Legend: Mar83 - \citet{Martin1983}; Sch85 - \citet{Schmidt1985}; Ber90 - \citet{Berriman1990};
	    Bri90 - \citet{Brindle1990}; Nan97 - \citet{Nandra1997}; Sto97 - \citet{Storchi1997};
	    Bar99 - \citet{Barth1999}; Sch01 - \citet{Schmid2001}; Wu01 - \citet{Wu2001}; Smi02 - \citet{Smith2002a}; 
	    Zha02 - \citet{Zhang2002}; Sch03 - \citet{Schmid2003}; Hic08 - \citet{Hicks2008}; 
	    Bat11 - \citet{Batcheldor2011}; Fis13 - \citet{Fischer2013} and Ris13 - \citet{Risaliti2013}.}
  \label{Table:Type1}
\end{table*}

\begin{table*}
  \centering
  {   
   \begin{tabular}{|c|c|c|c|c|c|c|c|}
   \hline {\bf Object}		& {\bf Waveband (\AA)}	& {\bf Pol. degree (\%)}	& {\bf Pol. angle ($^\circ$)}	& {\bf Ref.}	& {\bf Inclination ($^\circ$)}	& {\bf Ref.}	& {\bf Method}\\
   \hline 0019+0107		& 4000 -- 8600 		& $>$ 0.98 			& 35.0 $\pm$ 0.5		& Ogl99		& 90.0				& Bor10		& H\\
	  0145+0416		& 1960 -- 2260 		& $>$ 2.14 			& 126.0 $\pm$ 1.0		& Ogl99		& 80.0				& Bor10		& H\\
	  0226-1024		& 4000 -- 8600 		& $>$ 1.81 			& 167.1 $\pm$ 0.2		& Ogl99		& 87.0				& Bor10		& H\\
	  0842+3431		& 4000 -- 8600 		& $>$ 0.51 			& 27.1 $\pm$ 0.6		& Ogl99		& 78.0				& Bor10		& H\\
	  1235+1453		& 1600 -- 1840 		& $>$ 0.75 			& 175.0 $\pm$ 12.0		& Ogl99		& 76.0				& Bor10		& H\\
	  1333+2840		& 4000 -- 8600 		& $>$ 4.67 			& 161.5 $\pm$ 0.1		& Ogl99		& 80.0				& Bor10		& H\\
	  1413+1143		& 4000 -- 8600 		& $>$ 1.52 			& 55.7 $\pm$ 0.9		& Ogl99		& 88.0				& Bor10		& H\\
          Circinus		& 5650 -- 6800 		& 22.4 -- 25.0			& 45.0				& Ale00		& 65.0				& Fis13		& F\\
          IRAS~13349+2438	& 3200 -- 8320 		& 23 -- 35			& 124.0 $\pm$ 5.0		& Wil92		& 52.0				& Wil92		& I\\
	  Mrk~3			& 5000 			& 7.77 -- 8.61			& 167.0				& Tra95		& 85.0				& Fis13		& F\\
	  Mrk~34		& 3200 -- 6200 		& $>$ 3.92			& 53.0 $\pm$ 4.5		& Kay94		& 65.0				& Fis13		& F\\	
	  Mrk~78		& 3200 -- 6200 		& 21.0 $\pm$ 9.0		& 75.3 $\pm$ 11.2		& Kay94		& 60.0				& Fis13		& F\\	
	  Mrk~573		& 3200 -- 6200 		& $>$ 5.56			& 48.0 $\pm$ 2.0		& Kay94		& 60.0				& Fis13		& F\\	
	  Mrk~1066		& 3200 -- 6200 		& $>$ 1.99			& 135.1 $\pm$ 2.6		& Kay94		& 80.0				& Fis13		& F\\	
	  NGC~1068		& 3500 -- 5200 		& 16.0 $\pm$ 2.0		& 95.0				& Mill83	& 70.0				& Hon07		& G\\		  
	  NGC~1667		& 5100 -- 6100 		& 0.35 -- 9.8			& 94.0 $\pm$ 1.0		& Bar99		& 72.0				& Fis13		& F\\	
	  NGC~4507		& 5400 -- 5600 		& 14.8 -- 16.3			& 37.0 $\pm$ 2.0		& Mor00		& 47.0				& Fis13		& F\\	
	  NGC~5506		& 3200 -- 6200 		& $>$ 2.6			& 72.8 $\pm$ 4.5		& Kay94		& 80.0				& Fis13		& F\\	
	  NGC~5643		& 5000 -- 5900 		& $>$ 0.75			& 57.0 $\pm$ 9.0		& Bri90		& 65.0				& Fis13		& F\\	
	  NGC~7674		& 3200 -- 6200 		& 6.54 -- 7.6			& 31.0				& Tra95		& 60.0				& Fis13		& F\\	
   \hline
   \end{tabular}
  }
  \caption{Recorded average continuum polarization states and inclinations of 20 type-2 AGN.
	    The first reference column is related to polarization measurements, the second to estimations
	    of the orientation. Methods used to determine the inclination of the system are described in Sect.~\ref{Comp:Incl}.
	    Legend: Mill83 - \citet{Miller1983}; Bri90 - \citet{Brindle1990}; Wil92 - \citet{Wills1992};
	    Kay94 - \citet{Kay1994}; Tra95 - \citet{Tran1995a}; Bar99 - \citet{Barth1999}; Ogl99 - \citet{Ogle1999}; 
	    Ale00 - \citet{Alexander2000}; Mor00 - \citet{Moran2000}; Hon07 - \citet{Honig2007};
	    Bor10 - \citet{Borguet2010} and Fis13 - \citet{Fischer2013}.}
  \label{Table:Type2}
\end{table*}

\subsection{Inclination versus polarization}
\label{Comp:Survey}

\begin{figure}
   \centering
      \includegraphics[trim = 10mm 0mm 0mm 0mm, clip, width=9.2cm]{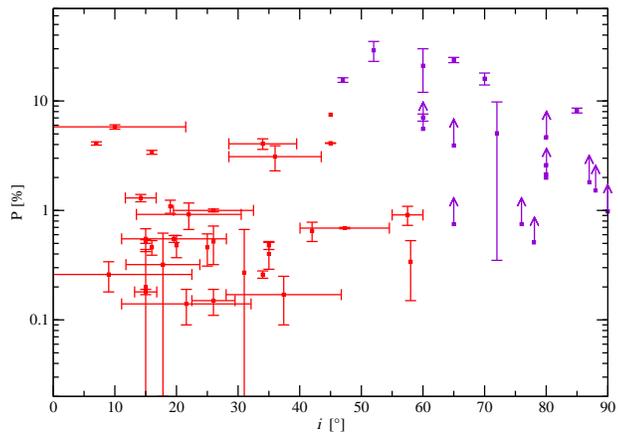}
      \caption{The polarization degree $P$ is plotted versus the 
	       AGN inclination $i$. Type-1 Seyfert-like galaxies are
	       shown in red, type-2 objects in violet.}
     \label{Fig:Survey}
   \end{figure}

The resulting compendium, comparing the polarization percentage $P$ versus the inclination $i$ for 53 objects, is presented in Fig.~\ref{Fig:Survey}. 
Type-1 AGN are shown in red, type-2 AGN in violet. 

According to the inclination estimations, type-1 AGN cover a range of orientation from $i$~=~0$^\circ$ (pole-on) to $i$~=~45$^\circ$ -- 60$^\circ$,
which is in agreement with the estimations of the overall half opening angle of the system, $\theta$~$>$~58$^\circ$, made by \citet{Osterbrock1993} 
and \citet{Ho1995}. Type-1 objects exhibit low polarization degrees ($\le$~1~\%), except for seven unusually, highly polarized sources (ESO~323-G077, 
Fairall~51, IC~4329A, MCG-6-30-15, Mrk~486, Mrk~766 and mrk~1239). Those high levels of polarization are similar to the detection of polarization degrees 
up to 4~\% in Mrk~231 \citep{Gallagher2005}, or in the Warm Infrared Ultraluminous AGN survey done by \citet{Hines1994}, but they still need to be 
explained (see Sect.~\ref{Discussion:Type1}). However, most of the type-1 sources collected here follow the empirical ascertainment, started with 
the observational surveys of Seyfert~1 AGN realized by \citet{Berriman1989}, \citet{Berriman1990} and \citet{Smith2002a}: type-1 AGN predominantly show low levels 
of polarization, associated with polarization position angles roughly parallel\footnote{Exceptions in the compendium are NGC~5548, Fairall~51, Mrk~486 
\citep{Smith2002a}, ESO~323-G077, NGC~3227, and probably Mrk~766 and NGC~4593 \citep{Batcheldor2011}, which exhibit polarization position angles 
perpendicular to their radio axis.} to the radio axis of the system (when radio axis measurements were available). It is particularly interesting 
to note that, among the seven polar scattering dominated AGN (with ``perpendicular'' polarization) identified, five exhibit a polarization degree 
higher than 1~\%. If a perpendicular polarization position angle can be explained by an increase of the opacity of the polar outflows (through which 
the observer's line of sight is passing), it is more difficult to explain such high $P$ at inclinations below 45$^\circ$ \citep{Marin2012a}.

In the case of type-2 AGN, objects show perpendicular polarization position angles associated with higher polarization percentages, not uncommonly 
$>$~7~\% (after first-order correction for the intrinsic continuum polarization). As electron-induced $P$ depends on the cosine square of the scattering 
angle, Seyfert galaxies seen edge-on are expected to be intrinsically more polarized \citep{Miller1983,Brindle1990,Kay1994}. Radiation scatters 
perpendicularly on to the ionized, polar outflows detected in type-2 AGN, leading to higher polarization percentages than at polar inclinations, 
where the combination of forward scattering and dilution by the unobscured nucleus diminishes the net polarization. A clear estimation of the average 
polarization percentage of Seyfert 2 objects remains problematic since most of the quoted values are lower limits. However, the trend is that $P$ seems 
to increase with inclination, with a possible decrease at extreme type-2 inclinations that has to be properly observed by future, rigorous measurements 
of the scattered light alone. The overall inclination of Seyfert~2 AGN lies between 60$^\circ$ and 90$^\circ$ (edge-on).

No clear polarization break is detected at the transition angles (45$^\circ$ -- 60$^\circ$) between type-1 and type-2 objects; $P$ increases 
continuously from pole-on to edge-on view. Note that the range of orientation that separates type-1 and type-2 Seyfert galaxies is not artificially 
enhanced by the selection criteria (see Sect.~\ref{Comp:Tables}), as none of the rejected inclinations cover the 45$^\circ$ -- 60$^\circ$ 
range. However, the reader is advised to note that possible bias in the estimation of $i$ may shift some objects to different inclination values. 
ESO~323-G077 (type-1), NGC~1365 (type-1), IRAS~13349+2438 (type-2) and the Circinus galaxy (ESO~97-G13, type-2) are in the intermediate zone 
between the two classifications. While the type-2 classification of the Circinus galaxy is undisputed, the cases of ESO~323-G077, NGC~1365 and 
IRAS~13349+2438 are more ambiguous. Spectropolarimetric observations achieved by \citet{Schmid2003} showed that, at least, a fraction of the inner 
region of ESO~323-G077 must be hidden behind the torus horizon, classifying this AGN as a borderline Seyfert~1 galaxy. NGC~1365 is another intriguing, 
borderline object, showing rapid transition between Compton-thin and Compton-thick regimes due to X-ray eclipses \citep{Risaliti2005}. Such behaviour 
leads to a difficult classification of NGC~1365, either type-1 \citep{Schulz1994}, type-1.5 \citep{Veron1980}, type-1.8 \citep{Alloin1981,Risaliti2005} 
or even type-2 \citep{Chun1982,Rush1993}. The case of IRAS~13349+2438 is somewhat similar to the previous object. The H$\beta$ line width measured 
by \citet{Lee2013} favours a type-2 classification, an argument in contradiction with \citet{Brandt1996} who found that IRAS~13349+2438 shares many 
properties with narrow-line Seyfert 1s, though they noted that this object must have a peculiar geometry. Such statement is supported by the 
spectropolarimetric observations realized by \citet{Wills1992}, who have shown that the observer's line of sight is probably intercepting part 
of the equatorial dusty material, classifying IRAS~13349+2438 as a borderline type-2 AGN. Thus, from a polarimetric point of view and if its 
estimated inclination is correct, the Circinus galaxy should be considered as a borderline Seyfert~2 object.

%%%%%%%%%%%%%%%%%%%%%%%%%%%%%%%%%%%%%%%%%%%%%%%%%%%%%%%%%%%%%%%%%%%%%%%
\section{Polarization predictions from theoretical models}
\label{Analysis}

A direct application of this polarization/inclination study concerns radiative transfer in numerical AGN models. While more recent simulations tend to 
complexify in terms of morphology, composition and kinematic, testing the relevance of a model against observations is a significant consistency check. 
In the following section, I run Monte Carlo simulations on four different AGN models from the literature and compare the polarimetric results to the 
compendium. 

Simulations of emission, multiple scattering and radiative coupling in complex AGN environments were achieved using {\sc stokes} 
\citep{Goosmann2007,Marin2012a}, a public\footnote{http://www.stokes-program.info/} Monte Carlo code including scattering-induced polarization. 
The results presented hereafter are representative of each different model, characterized by a unique set of parameters. The resulting wavelength-independent 
polarization percentage is integrated over 2000 to 8000~\AA. For consistency, the same input spectrum is used for all the models, namely an isotropic 
source emitting an unpolarized spectrum with a power-law SED $F_{\rm *}~\propto~\nu^{-\alpha}$ ($\alpha$~=~1). Finally, an important condition on the models 
investigated below is that, at least, they reproduce the expected polarization dichotomy (i.e. parallel polarization position angles for type-1 inclinations 
and perpendicular polarization position angles for type-2s).

\subsection{Three-component AGN}
\label{Analysis:3Comp}

According to the axisymmetric unified model \citep{Antonucci1993}, the central supermassive black hole and its accretion disc, that radiates 
most of its bolometric luminosity in the UV/optical bands, are obscured by an equatorial, dusty torus. A common hypothesis is that the
funnel of the torus collimates ejection winds in the form of a bi-conical polar outflow. Past spectropolarimetric models of AGN,
composed by a central irradiating source, a dusty torus and a bi-conical, electron-filled wind, showed that only perpendicular polarization
(with respect to the symmetry axis of the torus) can emerge \citep{Kartje1995,Marin2012a}. To introduce the production of parallel polarization 
in polar viewing angles, a third, equatorial region lying between the torus and the source has been proposed \citep{Antonucci1984,Young2000}. 
This highly ionized, geometrically thin disc can be associated with the accretion flow between the torus and the BLR 
\citep{Young2000,Goosmann2007} and is necessary to reproduce the observed polarization dichotomy \citep{Goosmann2007,Marin2012a}.

   \begin{figure}
   \centering
      \includegraphics[trim = 0mm 125mm 120mm 120mm, clip, width=8cm]{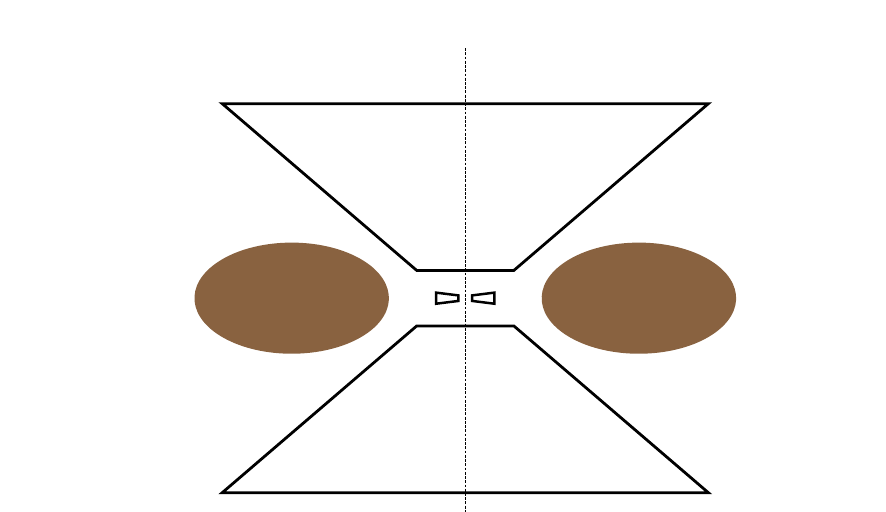}
      \includegraphics[trim = 10mm 0mm 0mm 8mm, clip, width=9.2cm]{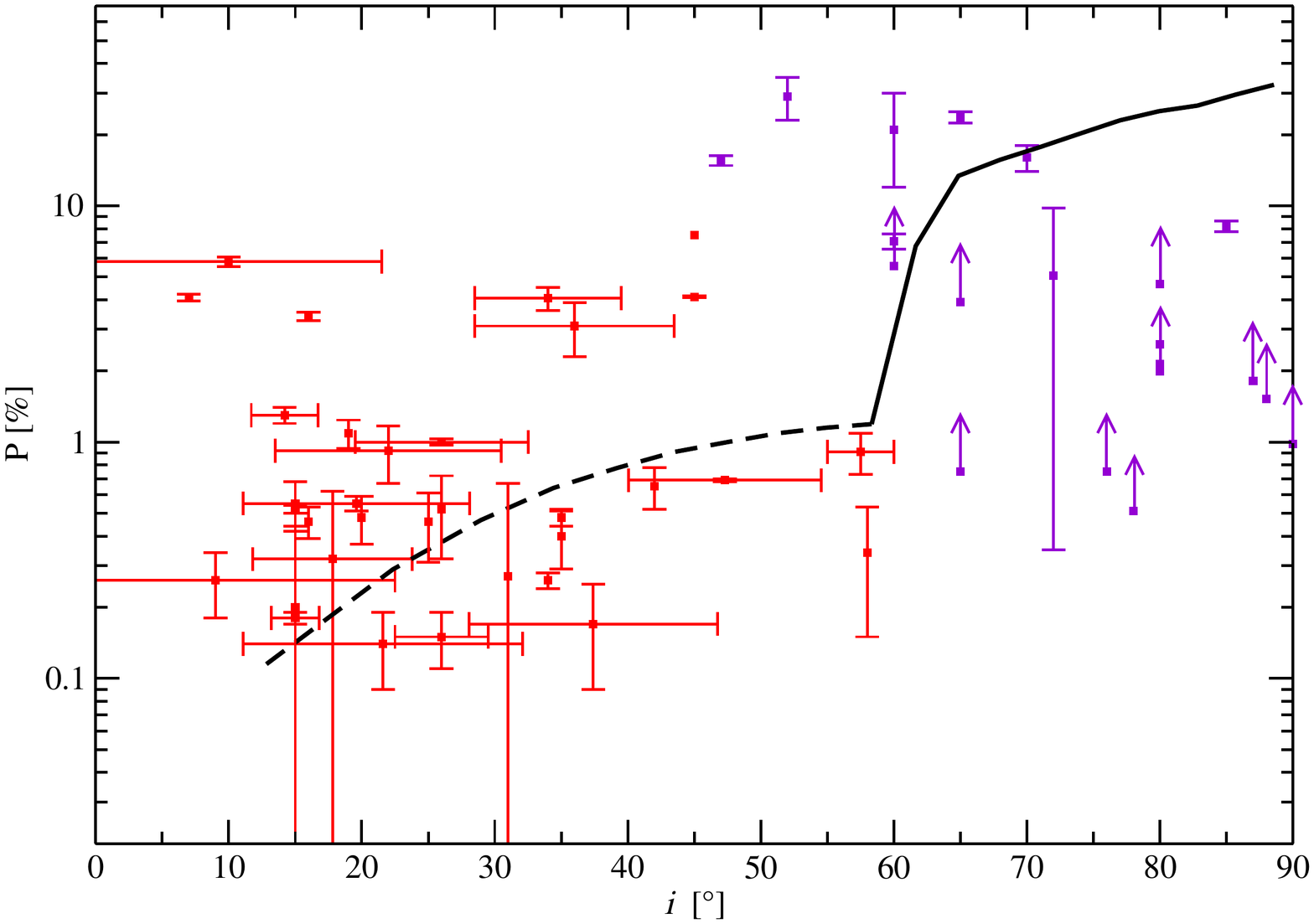}
      \caption{Top: schematic view of the three-component model. The dusty torus
	      is shown in dark brown.
	      Bottom: the resulting polarization (black line) of a three-component 
	      model (see \citealt{Marin2012a}) is plotted against observations.
	      The dashed section corresponds to parallel polarization, the solid line
	      to perpendicular polarization.}
     \label{Fig:3Comp}
   \end{figure}

Following the parametrization from \citet{Marin2012a}, the central, unpolarized source is surrounded by an equatorial, scattering flared disc 
with a half-opening angle of 20$^\circ$ with respect to the equatorial plane and a Thomson optical depth in the $V$ band of $\tau_{disc}$~=~1. 
Along the same plane, an optically thick ($\tau_{torus}~\gg$~1), dusty torus, filled with a standard ``Milky Way'' dust mixture \citep{Mathis1977} 
prevents radiation to escape along the equator. An hourglass-shaped, electron-filled region ($\tau_{wind}$~=~0.3) accounts for the polar 
ejection flow. The torus and the collimated ionized wind sustain the same half-opening angle, 60$^\circ$ with respect to the symmetry axis of the 
model, see Fig.~\ref{Fig:3Comp} (top). The reader may refer to \citet{Marin2012a} for further details about the model.

The wavelength-integrated polarization spectrum of the three-component model is shown in Fig.~\ref{Fig:3Comp} (bottom). From 0$^\circ$ to 60$^\circ$ 
(where the transition between type-1 and type-2 classification occurs), the three-component model successfully reproduces the average 
polarization level expected from type-1 AGN (i.e. $P~\le$~1~\%) as well as a parallel polarization position angle. $P$ rises from pole-on 
view to intermediate inclination without exceeding 1~\%. When the observer's line of sight crosses the torus height, $P$ decreases. This is 
due to the competition between parallel polarization produced by the equatorial, scattering disc and perpendicular polarization originating from the 
torus/polar regions, cancelling each other. Such behaviour is expected in any axisymmetric model and $P$ may decrease down to zero, depending 
on the global morphology of the system. Once the equatorial, electron-filled disc disappears behind the torus horizon, $P$ strongly increases, 
up to 30~\%, as radiation becomes dominated by perpendicular, Thomson scattering inside the polar outflows. The resulting polarization is not
high enough in the 45$^\circ$ -- 65$^\circ$ range and becomes too strong at large inclinations to fit the majority of type-2 objects.

\subsection{Four-component AGN}
\label{Analysis:4Comp}

\begin{figure}
   \centering
      \includegraphics[trim = 80mm 120mm 30mm 120mm, clip, width=8cm]{Schema.pdf}
      \includegraphics[trim = 10mm 0mm 0mm 8mm, clip, width=9.2cm]{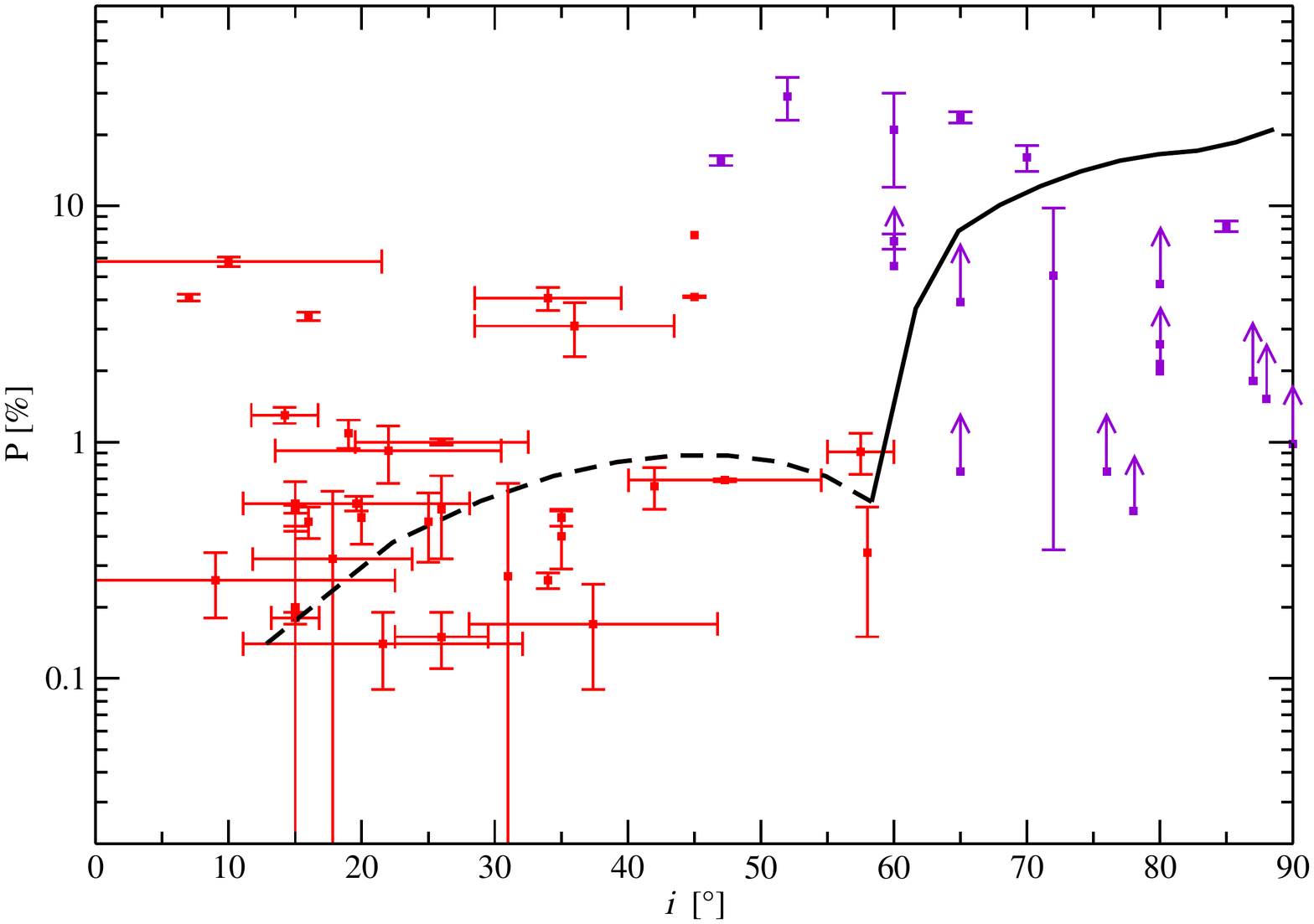}
      \caption{Top: schematic view of the four-component model. The model is the same as the 
	      three-component one with the addition of dusty NLR (shown in light brown).
	      Bottom: the resulting polarization (black line) of a four-component  
	      model (see \citealt{Marin2012b}) is plotted against observations.
	      The dashed section corresponds to parallel polarization, the solid line
	      to perpendicular polarization.}
     \label{Fig:4Comp}
   \end{figure}

A three-component model produces too much polarization at extreme type-2 inclinations. A natural way to decrease the amount of $P$ is to add an 
absorbing medium to the previous model. We know from observations that, beyond the dust sublimation radius, the ionized outflows merge continuously 
with the dusty environment of the host galaxy, forming the so-called (low-density) NLR \citep{Capetti1996,Capetti1999}. The next step is then to 
investigate, in the framework of this compendium, the polarization signature of a four-component model that includes a dust-filled, low opacity 
($\tau_{NLR}$~=~0.3) NLR \citep{Marin2012b}. I used the same parametrization as in Sect.~\ref{Analysis:3Comp}, with the NLR bi-cone sustaining 
the same half-opening angle as the ionized outflows (Fig.~\ref{Fig:4Comp}, top).

The addition of NLR into the modelling of an AGN does not strongly impact the overall polarization signature. It slightly decreases the net polarization in 
type-1 viewing angles (see Fig.~\ref{Fig:4Comp}, bottom) but does not alter its polarization position angle due to the large opening angle of the torus. 
However, the possibility to reach polarization degrees of few percents (such as for IC~4329A, MCG-6-30-15 or Mrk~766) becomes less likely due to absorption. 
The transition between parallel and perpendicular polarization occurs at the same inclination but the net polarization is lower, due to absorption. Finally, 
in comparison with observations, $P$ is found to be still too small between 45$^\circ < i <$ 65$^\circ$ ($P~\sim$~1 -- 10~\%).

\subsection{Fragmented media}
\label{Analysis:Uniform:Frag}

Optical and UV observations of the NLR of NGC~1068 \citep{Evans1991,Capetti1995a,Capetti1997,Packham1997} revealed the presence of many 
knots of different luminosity in the outflowing gas that can be attributed to inhomogeneities of the medium. Similar results are found for other sources
(e.g. Mrk~3; \citealt{Capetti1995b}), strengthening the idea that AGN outflows may not be a continuous flow \citep{Dai2008}. Due to the torus
compactness, there is less direct evidence for a clumpy torus and most of the suppositions about the fragmented nature of the circumnuclear matter 
comes from numerical simulations \citep{Pier1992,Pier1993,Nenkova2002}. 

\begin{figure}
   \centering
      \includegraphics[trim = 80mm 180mm 30mm 65mm, clip, width=8cm]{Schema.pdf}
      \includegraphics[trim = 10mm 0mm 0mm 8mm, clip, width=9.2cm]{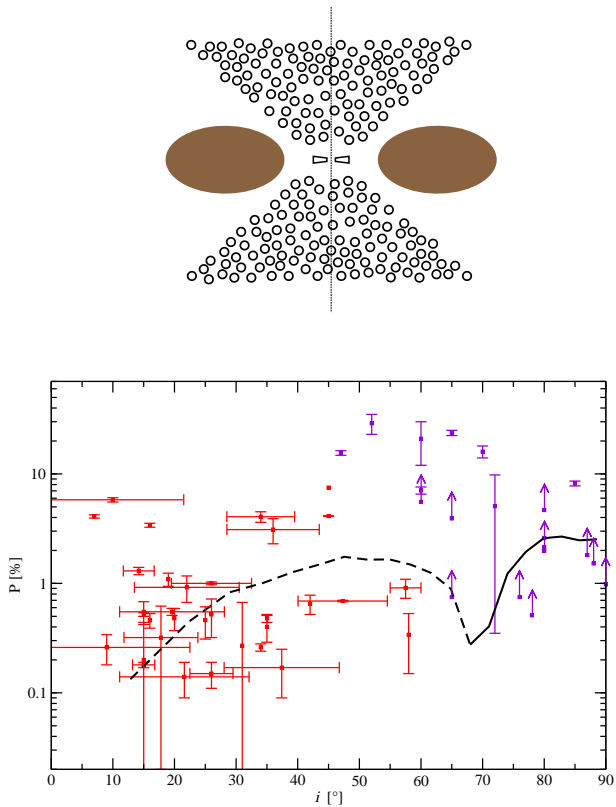}
      \caption{Top: schematic view of the clumpy model. The torus and the 
	      equatorial disc are the same as in Fig.~\ref{Fig:3Comp}.
	      Bottom: the resulting polarization (black line) of a  
	      clumpy model is plotted against observations.
	      The dashed section corresponds to parallel polarization, the solid line
	      to perpendicular polarization.}
     \label{Fig:Frag}
   \end{figure}

I now investigate a model in which the ionization cones are fragmented, while maintaining a compact dusty torus and an equatorial, 
scattering disc responsible for the production of parallel polarization. The equatorial disc and the circumnuclear dusty region retain the same 
morphological and composition parameters as in Sect.~\ref{Analysis:3Comp} and Sect.~\ref{Analysis:4Comp}. The fragmented outflows now consist of 
2000 electron-filled spheres of constant density ($\tau_{spheres}$~=~0.3; \citealt{Ogle2003}) and radius (filling factor $\sim$~5~\%). The 
filling factor is evaluated by summing up the volume of the clumps and dividing the total by the volume of the same unfragmented bi-conical NLR. 
A schematic view of the model is presented in Fig.~\ref{Fig:Frag} (top).

Similarly to the three and four-component models, a clumpy AGN (Fig.~\ref{Fig:Frag}, bottom) reproduces both the low polarization levels and the
expected polarization position angle in type-1 orientations. However, due to multiple scattering on the outflow's clumps increasing the polarization
degree, and gaps along the type-1 line of sights that allow a direct view of the electron disc, the net polarization percentage is higher, up to 2~\%
for intermediate inclinations. This level is still not sufficient to reproduce the observed polarization of highly polarized type-1 objects, but
a fragmented medium enables higher polarization degrees for type-1 modelling. Fragmentation also impacts the inclination at which perpendicular polarization 
starts to dominate the production of parallel polarization, but the resulting transition inclination is not consistent with the observed polarization
position angle of type-2 AGN. At type-2 inclinations, $P$ is much lower ($P~<$~3~\%) than in previous modellings due to the enhanced escape probability 
from the outflows. A fragmented model can match the lower limit on polarization of a large fraction of type-2 AGN but fails to reproduce the high 
continuum polarization of NGC~1068, Mrk~78, IRAS~13349+2438 or the Circinus galaxy. However, the cloudlet distribution is probably different for each 
individual object and should be adapted case by case, i.e. by increasing/decreasing the filling factor. By increasing the filling factor of the 
clumpy outflows, the model will start to behave like the AGN model presented in Sect.~\ref{Analysis:3Comp}, strengthening the production of 
perpendicular polarization at type-2 viewing angles and matching higher polarization percentages.

\subsection{A structure for quasars}
\label{Analysis:Elvis}

The hydrostatic equilibrium hypothesis, postulated for the equatorial, toroidal region, is slowly evolving. Based on the pioneering work done by 
\citet{Blandford1982}, a hydrodynamical scenario is now considered as an alternative to the usual dusty torus, involving clumps of dusty matter
embedded in a hydromagnetic disc-born wind (see \citealt{Elitzur2006}, and references therein). \citet{Elvis2000} took advantage of this scenario to
build a model which attempts to explain the broad and narrow absorption line regions, as well as the broad emission line region, of type-1 quasars. 
In its phenomenologically derived structure, a flow of warm, highly ionized matter (WHIM) arises from an accretion disc in a narrow range of 
radii, bent outward and driven into a radial direction by radiation pressure. The model of \citet{Elvis2000} was recently explored by 
\citet{Marin2013a,Marin2013b,Marin2013c}, who proposed a number of adjustments to match observed polarization data in the UV and optical bands. 

\begin{figure}
   \centering
      \includegraphics[trim = 0mm 180mm 120mm 70mm, clip, width=8cm]{Schema.pdf}
      \includegraphics[trim = 10mm 0mm 0mm 8mm, clip, width=9.2cm]{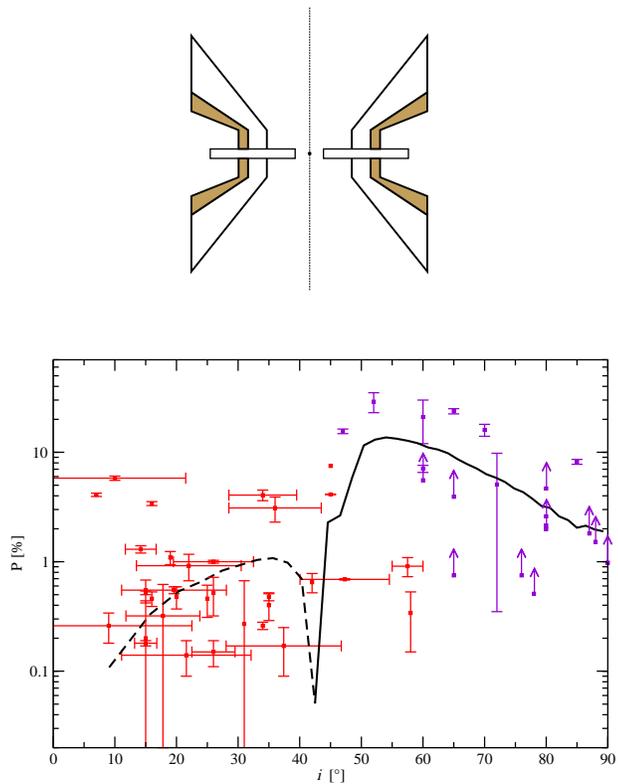}
      \caption{Top: schematic view of the structure for quasar as 
	      proposed by \citet{Elvis2000} and modified by \citet{Marin2013a}.
	      The WHIM appears in white, the failed dusty wind in brown.
	      Bottom: the resulting polarization (black line) of the  
	      disc-born wind model is plotted against observations.
	      The dashed section corresponds to parallel polarization, the solid line
	      to perpendicular polarization.}
     \label{Fig:Elvis}
   \end{figure}

To explore the consistency of the model described by \citet{Elvis2000} and modified by \citet{Marin2013a}, I plotted in Fig.~\ref{Fig:Elvis} (bottom)
the adjusted model proposed by \citet{Marin2013a}. The WHIM bending angle is set to 45$^\circ$ and its collimation angle to 3$^\circ$. It arises
at a distance $r$~=~0.0032~pc from the central source and extends up to 0.032~pc. The Thomson optical depth at the outflow's base and inside the 
conical, outflowing direction are, respectively, set to $\tau_{base}$~=~0.02 and $\tau_{flow}$~=~2. A failed wind, composed of cold dust, is 
self-shielded from the continuum source by the WHIM. The dusty outflow sustains a half-opening angle of 51$^\circ$ with respect to the symmetry 
axis of the system, and a collimation angle of 3$^\circ$. Its opacity along the equator is set to $\tau_{dust}$~=~4. Refer to \citet{Marin2013a}
for details about this choice of parameters.

From Fig.~\ref{Fig:Elvis} (bottom), it can be seen that the continuum polarization arising at type-1 inclinations follow the same trend as the three 
previous models: $P$ reaches a maximum value of 1~\% and cannot account for the highly polarized type-1 objects of the compendium. A local diminution
of $P$ appears at $i$~=~42$^\circ$, when the observer's line of sight crosses the outflowing material. The polarization position angle then switches 
from parallel to perpendicular, with respect to the projected symmetry axis of the system. While the transition between parallel and perpendicular
polarization occurs at a smaller $i$ in comparison with previous modelling, it is still coherent with the polarization position angle measurements of
NGC~5548 and ESO~323-G077, which exhibit perpendicular polarization at $i$~=~47.3${^\circ}^{+7.6}_{-6.9}$ and $i$~=~45$^\circ$, respectively. 
Beyond 54$^\circ$, when the observer's line of sight no longer passes through the radial outflows, $P$ reaches 10~\% -- 11~\% then slowly decreases 
because of the overwhelming impact of dust absorption along the equatorial direction. The polarization predicted by the line-driven wind model 
fits nearly all the observational, highly polarized type-2 objects in the 45$^\circ$ -- 65$^\circ$ range and can account for objects with
lower $P$ at extreme inclinations. It is noteworthy that while a disc-born wind is by far the closest model to observations, one must
be cautious as type-2 polarizations have first-order corrections and estimated inclinations are subject to potential biases.

%%%%%%%%%%%%%%%%%%%%%%%%%%%%%%%%%%%%%%%%%%%%%%%%%%%%%%%%%%%%%%%%%%%%%%%
\section{Discussion}
\label{Discussion}

\subsection{AGN modelling within the compendium}
\label{Discussion:Models}

The polarization-versus-inclination study presented in this paper allows a test of the relevance of four different AGN models from the literature.
All of them successfully reproduce both the observed polarization dichotomy and the average polarization percentage of type-1 AGN, but strongly differ 
in the 45$^\circ$ -- 90$^\circ$ inclination range. It is then easier to discriminate between several AGN models at intermediate inclinations. 
Models composed of uniform, homogeneous reprocessing regions (an equatorial scattering disc, a dusty torus and a pair of collimated cones, with the 
possible addition of dusty NLR) tend to create high perpendicular polarization degrees for type-2 AGN, while the polarization produced by a model 
with fragmented polar outflows do not extend farther than 3~\% in the same orientation range. Moreover, in the case of the model with clumpy ionization 
cones, the transition between parallel and perpendicular polarization happens at typical type-2 inclinations, which is in disagreement with observations. 
A refinement of the clumpy model is necessary. A deeper analysis of fragmented media is ongoing, targeting equatorial scattering discs, LIL and HIL BLR, 
tori, ionization cones and NLR. Preliminary results show that the polarization degree at type-2 inclinations can rise up to few tens of percent for dense 
cloudlet distributions. The transition between parallel and perpendicular polarization is correlated with the half-opening angle of the torus, and a 
fragmented circumnuclear region with a half-opening angle of 45$^\circ$ can produce a switch between parallel and perpendicular polarization position 
angle at $\sim$~60$^\circ$. The exploration of the parameter space of the models (optical depth, filling factor, covering factor ...) will be considered. 
The model of \citet{Elvis2000} is undoubtedly the closest to observation, as it produces both high and low polarization degrees at type-2 viewing angle, 
strengthening the hypothesis that at least some undermined fraction of AGN components are wind-like structures.

\subsection{Highly polarized type-1 AGN}
\label{Discussion:Type1}

Even by varying the parameters, none of the model can reach the high polarization levels of the inventoried type-1s ESO~323-G077 (7.5~\%), 
Fairall~51 (4.12~\% $\pm$ 0.03~\%), IC~4329A (5.80~\%~$\pm$~0.26~\%), MCG-6-30-15 (4.06~\%~$\pm$~0.45~\%), Mrk~486 (3.40~\%~$\pm$~0.14~\%), 
Mrk~766 (3.10~\%~$\pm$~0.80~\%), Mrk~1239 (4.09~\%~$\pm$~0.14~\%) or Mrk~231 ($\sim$~4~\%; \citealt{Gallagher2005}). Results presented in 
Sect.~\ref{Comp:Survey} show that their inclination ranges from 0$^\circ$ to 45$^\circ$, indicating that scattering at large angles between 
the photon source, the polar winds and the observer is unlikely to be responsible for all the atypically high continuum polarization observed. 
As the optical thickness of AGN polar outflows is estimated to be relatively small ($\tau~\ll$~1), the major contribution to scattering-induced 
type-1 polarization comes from the equatorial scattering disc. \citet{Goosmann2007} showed that the resulting polarization percentage from an 
equatorial, electron-filled disc is fairly low, independently of its half opening angle and Thomson opacity. A major challenge to numerical models 
of AGN is to increase the net polarization degree at type-1 viewing angles for isolated cases. It is even more challenging taking into account that 
the polarization position angle of five out of the seven objects (ESO~323-G077, Fairall~51, Mrk~486, Mrk~766 and NGC~3227) is found to be 
perpendicular to their radio axis, similarly to Mrk~231 \citep{Smith2004}.

There are several potential ways to strengthen the net polarization of the models. One can consider perpendicular scattering between the source, a 
reprocessing region located along the equatorial plane, and the observer. A promising target could be the HIL and LIL BLR. Considering the constraints on 
the HIL and LIL BLR structure derived by \citet{Kollatschny2013}, using kinematic measurements of emission lines in four nearby AGN, one can estimate the 
half-opening angle of the emission line region. According to \citet{Kollatschny2013}, the HILs (i.e. emitted close to the photoionizing source) 
originate from a medium with half-opening angle 3.8$^\circ$ -- 26$^\circ$ from the equatorial plane, while the LILs are created in a 
structure with a half-opening angle 11$^\circ$ -- 60$^\circ$. Further tests must be achieved to explore the polarization position angle and the amount 
of parallel polarization that HIL and LIL BLR can generate but it is unlikely that scattering within an axisymmetrical model can reach up to 
few percents at type-1 inclinations. 

As stated by \citet{Gaskell2010,Gaskell2011}, breaking the symmetric pattern of irradiation can help to understand the velocity dependence of broad 
emission line variability detected in many AGN. Strong off-axis flares could then explain the observed, extremely asymmetric, Balmer lines with broad peak 
redshifted or blueshifted by thousands of km$\cdot$s$^{-1}$ \citep{Smith2002a}, and produce higher polarization degrees even at polar orientation. It is 
important to test the off-axis flare model as, if correct, the azimuthal phase of the continuum source would play a critical role in the measurement of 
the supermassive black hole mass. \citet{Goosmann2013} have recently started the investigation of the off-axis irradiation theory by looking at the velocity 
dependence of the polarization of the broad emission lines. Preliminary results, compared to spectropolarimetric data for type-1 AGN from the literature, 
indicate that both the degree and position angle of polarization should be affected by asymmetrical emission. The net polarization percentage of the optical 
continuum is also slightly stronger. Further modelling will be achieved to explore how far optical and UV continuum polarization can be strengthened by 
temporary off-axis irradiation.

Finally, highly polarized AGN are not uncommon in type-1 radio-loud objects, with $P$ up to 45.5~\%~$\pm$~0.9~\% \citep{Mead1990}. The net polarization 
is quite variable from the radio to the optical band, often on short time-scales, pointing towards well-ordered magnetic fields surrounding a spatially 
small emitting region. While most of the observed polarization of radio-quiet AGN is undoubtedly originating from scattering off small particles 
\citep{Stockman1979,Antonucci1984,Antonucci1993}, the high, optical polarization of blazing quasars is thought to be associated with Doppler-boosted 
synchrotron emission from relativistic jets pointing towards us. The polarization position angle of blazar cores is usually perpendicular to the jet axis
while a parallel component is detected for emerging superluminal knots \citep{Darcangelo2009}. If electron and dust reprocessing appear to be unable to 
reach $P$~$\ge$~3~\% along poloidal directions, the correct interpretation might lie somewhere in the middle. The presence of a sub-parsec, aborted jet 
has yet to be proven but could explain the time variability and spectra of Narrow Line Seyfert~1 galaxies \citep{Ghisellini2004} and potentially create 
perpendicular polarization degree up to a few percent, while the lines and continuum polarization would be still mainly produced by reprocessing. If an 
undetermined fraction of the total polarization of highly polarized type-1 AGN indeed originates from synchrotron emission, the net polarization could be 
expected to vary, but with a much smaller amplitude than for synchrotron-dominated, radio-loud objects. Long term monitoring of radio-quiet, highly-polarized 
type-1 AGN would then help to evaluate the fraction of polarization arising from reprocessing and from synchrotron emission.

\subsection{Potential caveats on the determination of inclination}
\label{Discussion:Bias}

Estimations made by \citet{Wu2001} and \citet{Zhang2002} are primarily based on the assumptions that Seyfert~1 and normal galaxies follow the same black 
hole mass\footnote{To rectify black hole masses obtained from reverberation mapping, a correction factor $f$~=~5 was used by \citet{Ho1999}.} - bulge velocity 
dispersion correlation, and that the LIL BLR are in pure Keplerian rotation, coplanar to the system inclination. While detected, the motion and the morphology 
of the LIL BLR remain uncertain \citep{Peterson2006}. There are no strong constraints from the emission line profiles as a wide variety of kinematic 
models are able to reproduce the non-Gaussian profiles detected in AGN \citep{Bon2009}. The technique of velocity-resolved reverberation mapping 
\citep{Gaskell1988} is a step forward and tends to rule out any significant outflow from the AGN, while detecting a slight inflow \citep{Gaskell2007} 
and fast Keplerian motion. 

Fitting the distorted red wing of the Fe K$\alpha$ fluorescent line in X-ray bright, type-1 AGN can lead to possible bias. The procedure used by 
\citet{Nandra1997} assumes that the asymmetrical broadening of the iron line is caused by Doppler and general relativistic effects close to the central
black hole; however, a competitive mechanism was proposed by \citet{Inoue2003,Miller2008,Miller2009} and \citet{Miller2013}. In this scenario,
line broadening occurs at larger distances from the accretion disc, where a distribution of cold, absorbing gas blocks a fraction of the initial 
continuum. Transmitted and scattered radiation through the cloudlet environment finally carves out the distorted red wing. If distant absorption 
dominates relativistic effects, the estimated inclinations might then be questionable\footnote{In this context, a future X-ray polarimetric mission 
would be a solid tool to identify the preponderant mechanism responsible for line distortion \citep{Marin2012d,Marin2013d,Marin2013e}.}. 

The method developed by \citet{Fischer2013}, based on the work achieved by \citet{Crenshaw2000} and \citet{Das2005,Das2006}, relies on the nature of 
the NLR kinematics to determine the orientation of the system. One of their fundamental hypothesis is consistent with the unified model: AGN are
axisymmetrical objects. In this picture, the NLR structure sustains the same symmetry axis as the dusty torus, which is coplanar with the accretion 
disc. Thus, determining the inclination of the NLR is equivalent to determining the orientation of the whole system. However, a recent IR interferometric 
campaign carried out by \citet{Raban2009} found that the extended outflows of NGC~1068 are likely to be inclined by 18$^\circ$ with respect to 
the obscuring torus axis. If this trend is confirmed, and observed for other Seyfert-like galaxies, the overall AGN picture will become more complex.

Finally, the two-component model produced by \citet{Borguet2010} shows degeneracies between the various parameter combinations. It disallows the 
characterization of the outflow geometry in quasars showing broad absorption features, and weakens the constraints brought on the inclination of 
the system. The viewing angles of the outflow derived by the authors are quite large, a conclusion shared by \citet{Schmidt1999} and \citet{Ogle1999}, 
based on optical polarization surveys of BAL quasars, but contested by several other authors \citep{Barvainis1997,Punsly2010}. 
In particular, \citet{Punsly2010} showed that two-thirds of the BAL quasars they observed using H$\beta$ line width as a diagnostic are well
represented by objects with gas flowing along polar directions.

%%%%%%%%%%%%%%%%%%%%%%%%%%%%%%%%%%%%%%%%%%%%%%%%%%%%%%%%%%%%%%%%%%%%%%%
\section{Conclusions}
\label{Conclusion}

The first match of 53 AGN inclinations with their intrinsic continuum polarization originating from electron and dust scattering is presented 
in this paper. Different techniques to retrieve the nuclear orientation of type-1 and type-2 Seyfert galaxies were presented and discussed, 
highlighting their potential caveats. The continuum polarization of several Seyfert-2s was corrected using broad H$\alpha$ and H$\beta$ line 
polarization as a reliable indicator of the true polarization of the scattered light, and lower limits were put for the remaining AGN whose 
polarization spectra were either noise-saturated or unpublished. 

~\

The resulting compendium\footnote{The compendium will be regularly updated and available upon email request.} is in agreement with past 
observational/theoretical literature, and warrants additional conclusions and remarks the following.
\begin{itemize}
  \item Seyfert 1 AGN are associated with low polarization degrees, $P~\le$~1~\%, and predominantly characterized by a polarization position 
  angle parallel to the projected radio axis of the system. The inclination of type-1 objects ranges from 0$^\circ$ to 60$^\circ$.
  
  \item Seven type-1s have been identified as polar scattering dominated AGN, i.e. showing a perpendicular polarization position angle. 
  Among them, five have an atypical continuum polarization higher than 1~\%, mostly associated with 10$^\circ$ -- 45$^\circ$ inclinations.
  As scattering-induced polarization is unlikely to produce such high polarization degrees at type-1 orientation, a more elaborate scenario 
  must be considered. 
    
  \item After correction, Seyfert 2 AGN show polarization degrees higher than 7~\% and perpendicular polarization position angle.
  Unfortunately, most of the objects have only lower limits. The inclination of type-2 objects is ranging from 47$^\circ$ to 90$^\circ$.
  
  \item The transition between type-1 and type-2 AGN occurs between 45$^\circ$ -- 60$^\circ$. This range of inclination is likely to include 
  AGN classified as borderline objects, where the observer's line of sight crosses the horizon of the equatorial dusty medium. Four 
  objects lie in this range, three of them (ESO~323-G077, NGC~1365 and IRAS~13349+2438) being already considered as borderline Seyfert 
  galaxies from spectroscopic observations. If the estimated inclination of the fourth object (the Circinus galaxy) is correct, it 
  should be considered as another borderline AGN.
  
  \item The usual axisymmetric AGN models have difficulties to reproduce the trend of polarization with inclination. Fragmenting
  the reprocessing regions is helpful to cover a wide range of continuum polarization but a disc-born wind model is found to be 
  already quite close from observations.A fine tuning of the line-driven disc wind could easily  match a substantial fraction of the 
  reported measurements.
\end{itemize}

~\

Problems determining the inclination of AGN must be taken into consideration but, despite potential caveats, the associated continuum 
polarization lies within the margins of past empirical results, consolidating the basis of the compendium. It is then important for 
future models, as a consistency check, to reproduce the average continuum polarization, the polarization dichotomy and the transition 
between type-1 and type-2 classification (45$^\circ < i <$ 60$^\circ$). By improving the quality of the methods to determine the inclination 
of AGN and properly removing the contribution of both stellar and starburst light in future polarimetric measurement of type-2 objects,
it will possible to bring very strong constraints on the morphology, composition and kinematics of AGN. To achieve this goal, a new
UV/optical spectropolarimetric atlas of Seyfert 2s is necessary.

%%%%%%%%%%%%%%%%%%%%%%%%%%%%%%%%%%%%%%%%%%%%%%%%%%%%%%%%%%%%%%%%%%%%%%%
\section*{Acknowledgements}

The author is grateful to the referee Ski Antonucci for his useful and constructive comments on the manuscript. I also acknowledge
the Academy of Sciences of the Czech Republic for its hospitality, and the French grant ANR-11-JS56-013-01 of the project POLIOPTIX 
and the COST Action MP1104 for financial support.

%%%%%%%%%%%%%%%%%%%%%%%%%%%%%%%%%%%%%%%%%%%%%%%%%%%%%%%%%%%%%%%%%%%%%%%

\appendix

\bsp

\label{lastpage}

\end{document}